\documentclass[draft,jgrga]{agutex}
\def\degree{\hbox{$^\circ$}}

\usepackage{lineno}

\usepackage{graphicx}
\usepackage{leftidx}
\usepackage{color}

\setkeys{Gin}{draft=false}

\authorrunninghead{WANG ET AL.}

\titlerunninghead{Testing reanalysis datasets in Antarctica}

\authoraddr{Corresponding author: Dong Zhou, Department of Physics, Bar-Ilan University, Ramat-Gan 52900, Israel. (zhoudongbnu@gmail.com)}

\begin{document}


\title{Testing reanalysis datasets in Antarctica: Trends, persistence properties and trend significance}




\authors{Yang Wang,\altaffilmark{1}
Dong Zhou,\altaffilmark{1}, 
Armin Bunde, \altaffilmark{2}
Shlomo Havlin\altaffilmark{1}}

\altaffiltext{1}{Department of Physics, Bar-Ilan University, Ramat-Gan 52900, Israel}

\altaffiltext{2}{Institut f\"{u}r Theoretische Physik, Justus-Liebig-Universit\"{a}t Giessen, 35392 Giessen, Germany}



\begin{abstract}

{The reanalysis data sets provide important sources for investigating the climate in Antarctica where stations are sparse. In this paper, we compare the 2-meter-near-surface temperature data from 5 major reanalysis data sets with observational Antarctic stations data over the last 36 years: (i) the National Centers for Environmental Prediction and the National Center for Atmospheric Research Reanalysis (NCEP1), (ii) NCEP-DOE Reanalysis 2 (NCEP2),  (iii) the European Centre for Medium-Range Weather Forecasts Interim Reanalysis (ERA-Interim), (iv) the Japanese 55-year Reanalysis (JRA-55), and (v) the National Aeronautics and Space Administration (NASA) Modern-Era Retrospective-analysis for Research and Applications (MERRA). In our assessment, we compare (a) the  annual and seasonal trends obtained by linear regression analysis, (b) the {standard deviation} around the annual trends, (c) the detrended lag-1-autocorrelation $C(1)$, (d) the Hurst exponent $\alpha$ that characterizes the long-term memory in a record, and (e) the significance levels of the warming/cooling trends.  We find that all 5 reanalysis data sets are able to reproduce quite well the long-term memory in the instrumental data. In contrast, $C(1)$, which is needed as input for the conventional significance analysis shows fully erratic behavior. The observational warming/cooling trends in East  and West Antarctica are not reproduced well by all reanalysis data sets, in particular NCEP1, NCEP2, and JRA-55  show spurious warming trends in many parts of East Antarctica, even in those parts where cooling has been observed. In contrast, the standard deviation around the trends is quite well reproduced by all reanalysis data sets. In the Peninsula where the station density is quite high, the performance of the reanalysis data is considerably better. It is remarkable that all reanalysis data sets as well as the observational data show (under the assumption of a long-term persistent process) that  in the considered time period since 1979 the warming in the Peninsula is not significant with  $p$ values well above 0.1.}

\end{abstract}



\begin{article}

\section{Introduction}

In recent years, the warming patterns of Antarctica have received much attention {\citep[e.g.,][]{Turner2005,Steig2009,Ding2011,Bromwich2013}}. The large Antarctic Ice Sheet is one of the crucial tipping elements in the global climate system~\citep{Lenton12022008}, and a change of the dynamics and mass balance of the Antarctic Ice Sheet may have widespread implications, contributing and finally dominating the global and regional sea level rise~\citep{Turner2011}. Thus, addressing the Antarctic temperature change in a long-term context and determining its significance is an important issue~\citep{Schneider2006,Steig2013, Bromwich2013, Bromwich2014,Bunde2014,Ludescher2015,Tamazian2015}. However, this is quite a challenging problem due to the lack of long observational data sets.

Accordingly, reanalysis datasets  that are able to fill the gaps in data-sparse regions and help give the full physical picture of the climate change in Antarctica are highly needed and appreciated~\citep{Ding2011,Ding2012,Bromwich2013}. Reanalysis data also play a central role in reconstructing Antarctic near surface temperature~\citep{Monaghan2008,Steig2009,Nicolas2014}. The most commonly used physical quantities are mean sea level pressure (MSLP), geopotential height at 500-hPa ($Z500$), near surface air temperature (at 2 meter, $T_{2m}$), as well as geopotential height at higher pressure levels such as 30-hPa ($Z30$)~\citep{Thompson2002,Ding2011,Ding2012,Bromwich2013}.

Until now, several reanalysis datasets have been released. The most commonly used datasets include the National Centers for Environmental Prediction (NCEP) and the National Center for Atmospheric Research (NCAR) reanalysis (hereafter, NCEP1), NCEP-DOE Reanalysis 2 (hereafter, NCEP2), the European Centre for Medium-Range Weather Forecasts (ECMWF) Interim Re-Analysis (hereafter, ERA-Interim), the Japanese 55-year Reanalysis (hereafter, JRA-55), and the National Aeronautics and Space Administration (NASA) Modern-Era Retrospective-analysis for Research and Applications Reanalysis (hereafter, MERRA). Besides Antarctic research, these datasets are also widely used to study atmospheric dynamics and provide invaluable data sources for interdisciplinary research ~\citep[e.g.,][]{Rozanov2005GRL,Rodger2008GRL,Seppaala2013,Lam2015,Regi2016, Gozolchiani2011,Yamasaki2008, Donges2009, Berezin2012, Wang2013,Ludescher2013,Ludescher2014}. Accordingly, the reliability of these reanalysis data sets is of great interest.

{In this paper, we are interested in the quality of the near-surface (2m) temperature data provided by the 5 reanalysis data sets  over Antarctica. Previous evaluations of the performance of reanalysis data~\citep{Hines2000, Marshall2000, Marshall2003,Bromwich2004,Bromwich2007,Bromwich2011,Cullather2011,Bracegirdle2012,Jones2014,Nicolas2014} concentrated mainly on the biases of the data and the temperature patterns. Here we assess the reanalysis skills  by  focusing for the first time on (a) the annual and seasonal trends obtained by linear regression, (b) the {standard deviation} around the annual trends, (c) the detrended lag-1-autocorrelation $C(1)$, (d) the Hurst exponent $\alpha$ that characterizes the long-term memory in a record, and (e)  the significance levels of the annual warming/cooling trends.  We concentrate on the time span between January 1979 and December 2014, when the modern satellite data were assimilated.}

{It is obvious that the magnitude of a trend and the fluctuations around the trend line are central quantities of a temperature record, in particular in the context of  global
change. For describing the persistence of a record, $C(1)$ and $\alpha$ are central quantities. In most previous attempts to obtain quantitatively the significance of a warming or cooling trend it has been assumed that the data are short term persistent such that the detrended lag-$s$-autocorrelation function $C(s)$ decays exponentially, $C(s)=C(1)^{-s}$. As far as we know, there are no theoretical or observational  arguments why atmospheric (or sea surface) temperature should decay exponentially. The exponential decay has been assumed as the simplest way of describing the persistence of the record. Furthermore, there are no theories that describe the value of $C(1)$ as a function of the location of the considered station.
Since the persistence is fully determined by $C(1)$ in this case, the significance of a trend  is also determined solely by $C(1)$~\citep{Santer2000}, 
making $C(1)$ an essential quantity. }

In the past few decades it has been realized based on data analysis, that temperature records are not short-term persistent but long-term persistent all over the globe {~\citep[e.g.,][]{Hurst1951, Mandelbrot1968, Bloomfield1992, Koscielny1998, Malamud1999, Fraedrich2003, Eichner2003, Monetti2003, Vyushin2004, Cohn2005, Kiraly2006, Rybski2008, Lennartz2009, Franzke2010, Franzke2012,Lovejoybuch, Ludescher2015,Yuan2015}}. In long-term persistent records, $C(s)$ has been found not to decay exponentially, but by a power law, $C(s) \propto s^{-\gamma}$. The exponent  $\gamma$ can be obtained best by using the Detrended Fluctuation Analysis (DFA)~\citep{Peng1994,Kantelhardt2001}  (see Methods Section), where the Hurst exponent $\alpha=1-\gamma/2$ is measured. Typically, for coast line stations $\alpha$ is around 0.65 ($\pm 0.10$), while for island stations and sea surface temperature $\alpha$ is significantly larger, being around 0.75 ($\pm 0.15$) for island stations and around 0.8 ($\pm\ 0.10$) for sea surface temperatures~\citep{Koscielny1998, Malamud1999, Fraedrich2003,  Eichner2003,  Monetti2003,  Kiraly2006,  Franzke2010, Franzke2012,Ludescher2015,Yuan2015}. The error bars refer to the 95 percent confidence interval. Recently, it has been demonstrated explicitly by using DFA2 that the Antarctic temperature records cannot be considered as short-term persistent \citep{Bunde2014, Ludescher2015, Yuan2015}. Instead, as for the rest of the globe, also the Antarctic temperature data are long-term persistent, with $\alpha$ values similar to those found in other locations around the globe.

{There is no comprehensive theory that describes the origin of the long-term persistence in temperature data. However,  studies of  general circulation models (AOGCMs)  show that apart from the inertia of the oceans, the natural forcings play an important role, in particular volcanic forcing~\citep{Vyushin2004}. Since the atmospheric and the sea surface temperatures are long-term persistent, the Hurst exponent  $\alpha$ is the central quantity that quantifies the temperature `landscape' of a record,  and it is important to know if the reanalysis data are able to reproduce proper Hurst exponents or not. From the Hurst exponent of a long-term persistent record one can derive quantitatively the statistical significance of a warming or cooling trend \citep{Lennartz2009,  Lennartz2011, Tamazian2015}. Since in the context of global change the statistical significance of a trend is of great interest, we also compare their values for the different reanalysis data sets in Antarctica.}

{In this article, we are not the first to study  the temperature patterns in Antarctica and the long-term persistence in the near-surface temperature data.
We do not aim here to establish a theory for the various temperature patterns as well as for the long-term persistence with the slightly varying Hurst exponents.
 Instead, we wish to quantify the cooling/warming trends and the Hurst exponents in the observational and the reanalysis data and compare them to each other. This comprehensive comparison is important since the observational data is sparse in  Antarctica, and we need to know to which extent we can  rely on the reanalysis data sets, in order to investigate the Antarctic temperature change as well as the teleconnections with other regions in the globe. Our study shows that all reanalysis projects are unable to describe satisfactorily the annual and seasonal trends in West and East Antarctica where the station data are sparse. Regarding the dynamical behavior we find that $C(1)$ fluctuates very strongly in all reanalysis data sets, such that different reanalysis sets would provide very different persistent properties for the same station, ranging from intermediate anti-persistence to strong persistence in case the conventional picture of a dominating short-term persistent process was correct. We show that this is another indication that the Antarctic temperatures are not mainly short-term persistent.}

{Indeed, one of the remarkable results of this study is that all reanalysis data are able to reproduce the long-term persistence of the observational temperature records, with comparable Hurst exponents. The second remarkable result is that all reanalysis data sets except JRA-55 describe the situation in the Peninsula (which is considered to be one of the fast warming regions on Earth) reasonably well. The results show that from 1979 on the warming is not significant, with $p$ values well above 0.1. Accordingly, after 1979  the temperature trends  in the Peninsula are well within the bounds of natural variability, in agreement with the results of {\citet{Ludescher2015} and \citet{Turner2016}}.}

This paper is organized as follows. In section {2}, we provide the source of the Antarctic stations data and describe the five reanalysis data sets.  In section {3}, we present the results of our assessment. Section {4} concludes the article with a short summary. The Method Section 5 describes the methods, including long- and short-term persistence, detrended fluctuation analysis (DFA2),  and statistical significance analysis in detail.

\begin{figure*}[htpb]
    \centering
\includegraphics[width=\textwidth,trim=50 50 50 50,clip=false]{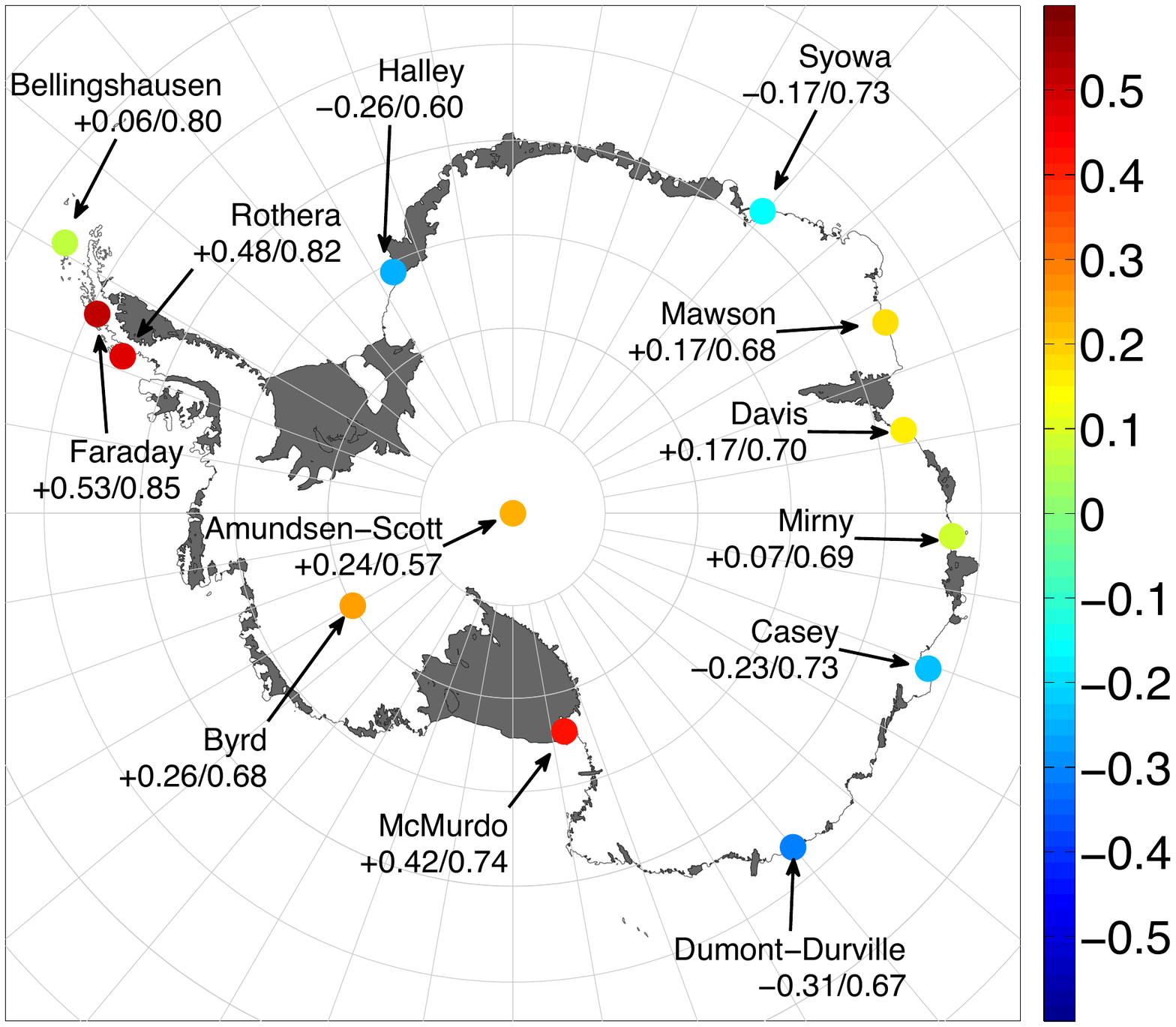}
  \caption{\textbf{The locations of the stations.} The linear regression analysis ($\Delta$) of temperature change from January 1979 to December 2014 and the Detrended Fluctuation Analysis (DFA2) (Hurst exponent $\alpha$) are shown ($\Delta/\alpha$) The color represents $\Delta$ with the units degrees of Celsius  per decade.}
  \label{Fig1}
\end{figure*}


\section{Data}
\subsection{Observation dataset}

In this study, we use monthly mean observation {$T_{2m}$ data (from January 1979 until December 2014) from 12 Antarctic stations (4 in West Antarctica including the Peninsula and 8 in East Antarctica) provided by the British Antarctic Survey Reference Antarctic Data for Environmental Research project (READER)~\citep{Turner2004a} (see http://www.antarctica.ac.uk/met/READER/). {The stations in the Peninsula are Bellingshausen, Rothera, and Faraday/Vernadsky, the West Antarctic station is McMurdo, and the East-Antarctic stations are Halley, Syowa, Mawson, Davis, Dumont-Durville, Mirny, and Amundsen-Scott (the South Pole). We have chosen these stations because they provide us with the longest reliable temperature records and also have been subject of a recent study on the persistence of Antarctic temperatures and the significance of their warming/cooling trends~\citep{Ludescher2015}. The READER dataset provides continuous observational temperature records for the Antarctic region.}  To fill the short temporal gaps in the data we apply linear interpolation. In addition, because of the lack of long records of station data in the interior of the West Antarctica, we also analyzed the reconstructed data set at Byrd station~\citep{Bromwich2013}. {Figure~\ref{Fig1} shows the  locations of all stations  studied here. At each location, we show the temperature change and the Hurst exponent that characterizes the long-term persistence in the considered time window.}

\subsection{Reanalysis data}

We consider five reanalysis datasets NCEP1, NCEP2, ERA-Interim, JRA-55 and MERRA, which are widely analyzed. NCEP1 uses a frozen state-of-the-art global assimilation system~\citep{Kalnay1996}. The data assimilation and the model used are identical to the global system implemented operationally at NCEP on January 11 1995, but with a reduced horizontal resolution of T62 ($\sim$ 209 \rm{km}). The spectral model used in the assimilation system contains 28 vertical levels from 1000hPa to 3hPa. Its temporal coverage is four times per day from January 1 1948 to the present day.

NCEP2 is an improved version of the NCEP1 model with the same horizontal and vertical resolutions, but fixed errors and updated parameterizations of the physical processes, such as the new boundary layer, new short wave radiation in the model and improved sea-ice SST fields~\citep{Kanamitsu2002}. The data records range from January 1 1979 to the present day. Both NCEP1 and NCEP2 reanalysis use a three-dimensional variational analysis scheme in the data analysis module. {The major input observational data for NCEP1/NCEP2 are the NCEP global upper air Global Telecommunication System (GTS) data  from 1962 provided by the National Center for Atmospheric Research (NCAR). Another data source is the NCEP global surface GTS data from 1967. Both data include all available Antarctic stations. These input  data are combined with other datasets, such as satellite data, surface marine data, surface wind data and so on. Therefore, the output $T_{2m}$ is strongly affected by both the observational data and the model parameterizations~\citep{Kalnay1996, Hines2000}.}

ERA-Interim, on the other hand, uses a four-dimensional variational assimilation scheme with a twelve-hour analysis window, and the data assimilation system is based on a 2006 version of the ECMWF Integrated Forecast Model. The spatial resolution of the model is T255 ($\sim$ 80 \rm{km}) on 60 vertical levels from the surface up to 0.1 hPa~\citep{Dee2011}. The data is continuously updated in real time from January 1 1979. Moreover, ERA-Interim and NCEP1/NCEP2 use different assimilation schemes of satellite data. While ERA-Interim assimilates raw satellite radiances, both NCEP1 and NCEP2 use satellite retrievals. Retrievals estimate the vertical temperature and humidity profiles through a series of empirical and statistical relationships, while raw radiances are direct measurements of atmospheric radiation acquired by the satellite sensors~\citep{Bromwich2007}. {Moreover, ERA-Interim reanalysis data uses the  GTS  surface  data  from  both  ECMWF and  external institutions, such as NCEP/NCAR and Japan Meteorological Agency (JMA).}

JRA-55 is known as the second reanalysis project conducted by the JMA, which uses a constant state-of-the-art but more complicated data assimilation system compared with their first reanalysis data set. {JRA-55 reanalysis mainly uses observational data provided by ECMWF, which includes both ECMWF and NCEP/NCAR surface and upper air observations from 1958. Therefore, it includes similar Antarctic stations as ERA-interim reanalysis. In addition, newly available observational datasets are collected whenever possible, including those used in past operational systems and delayed observations as well as digitized observations. High-quality reprocessed satellite data are also assimilated where available. This reanalysis data covers date from the year 1958, when regular radiosonde observation began on a global basis~\citep{Kobayashi2015}.}

MERRA is a NASA atmospheric data reanalysis for the satellite era using a major new version of the Goddard Earth Observing System Data Assimilation System Version five (GEOS-5). MERRA focuses on historical analyses of the hydrological cycle on a broad range of weather and climate time scales. MERRA also has very high spatial resolution (0.5\degree $\times$ 0.67\degree{}) that might show improved skills~\citep{Rienecker2011}. {MERRA reanalysis dataset mainly uses NCEP land surface observations from 1970. Therefore, all available Antarctic stations are also included as a part of input data. MERRA also used radiosonde data that were quality controlled by NCEP, with additional corrections. The conventional observational data used in MERRA include British Antarctic Survey radiosonde observational data.}

Here, we study the monthly mean $T_{2m}$ records in these five reanalysis datasets. {To compare with the observational station records we choose in the reanalysis records the nearest land point to a given selected station, since most stations analyzed are located near the Antarctic coast. However, if the nearest land point is more than 150~\rm{km} away (only in the Antarctic Peninsula), we choose the nearest reanalysis grid point instead. Note that many previous studies used the dry adiabatic lapse rate (9.8 \degree{C}$\rm{km^{-1}}$ or 6 \degree{C}$\rm{km^{-1}}$) to account for the Antarctic near-surface air temperature~\citep{Jones2014, Bracegirdle2012}. We do not need to use height adjustment in this study since all  quantities we are interested in  are independent of this correction.  }


\section{Results}
\subsection{Annual trends and standard deviation around the trend lines}

For obtaining the magnitudes of warming/cooling trends, we consider the annual near surface temperatures $T_i$, $i=1,2,\cdots,L=36$. We use linear regression analysis based on the least squares method, which minimizes the variances of $T_i$ along the regression line

\begin{equation}
	\hat{T}_i = \eta+bi.
\end{equation}

Accordingly, we define the total temperature change by $\Delta=b(L-1)$. The standard deviation around the regression line is $\sigma = \sqrt{\frac{1}{L} \sum\limits_{i=1}^{L} (T_i-\hat{T}_i)^2}$. The relative trend $x$ relevant for the statistical significance analysis is defined by the ratio between the total trend and the standard deviation, $x = \Delta/\sigma$, as defined by {\citet{Tamazian2015}}.

{Figure~\ref{Fig2} compares, for 4 of the 13 stations (Dumont-Durville, Amundsen Scott, Rothera, and McMurdo), the annual READER temperature time series (upper curves) with the corresponding data sets of the 5 reanalysis projects considered here. The straight lines are the regression lines. The y-axis is in degree Celsius. To avoid overlapping of the data, we have added to the different reanalysis sets arbitrary offsets (for transparency since we are interested in the trends and the structure of the temperature landscapes). One can see that the observational temperature trends are not reproduced well by all reanalysis data. For Dumont-Durville, the observed temperature trend is negative, but two of the data sets (NCEP1 and NCEP2) show a remarkable warming trend. For Amundsen-Scott, there is a moderate warming trend in the observational data. The reanalysis data show a mixed behavior, ranging from strong warming (NCEP1) to considerable cooling (ERA-Interim). In contrast, the variation of the data around the trend line is comparable in all cases, and also the characteristic `mountain-valley' structure of the READER data that is generated by their persistence properties, is roughly in line with the reanalysis data sets.}

\begin{figure}[htpb]
    \centering
\includegraphics[width=\columnwidth,trim=50 50 50 50,clip=false]{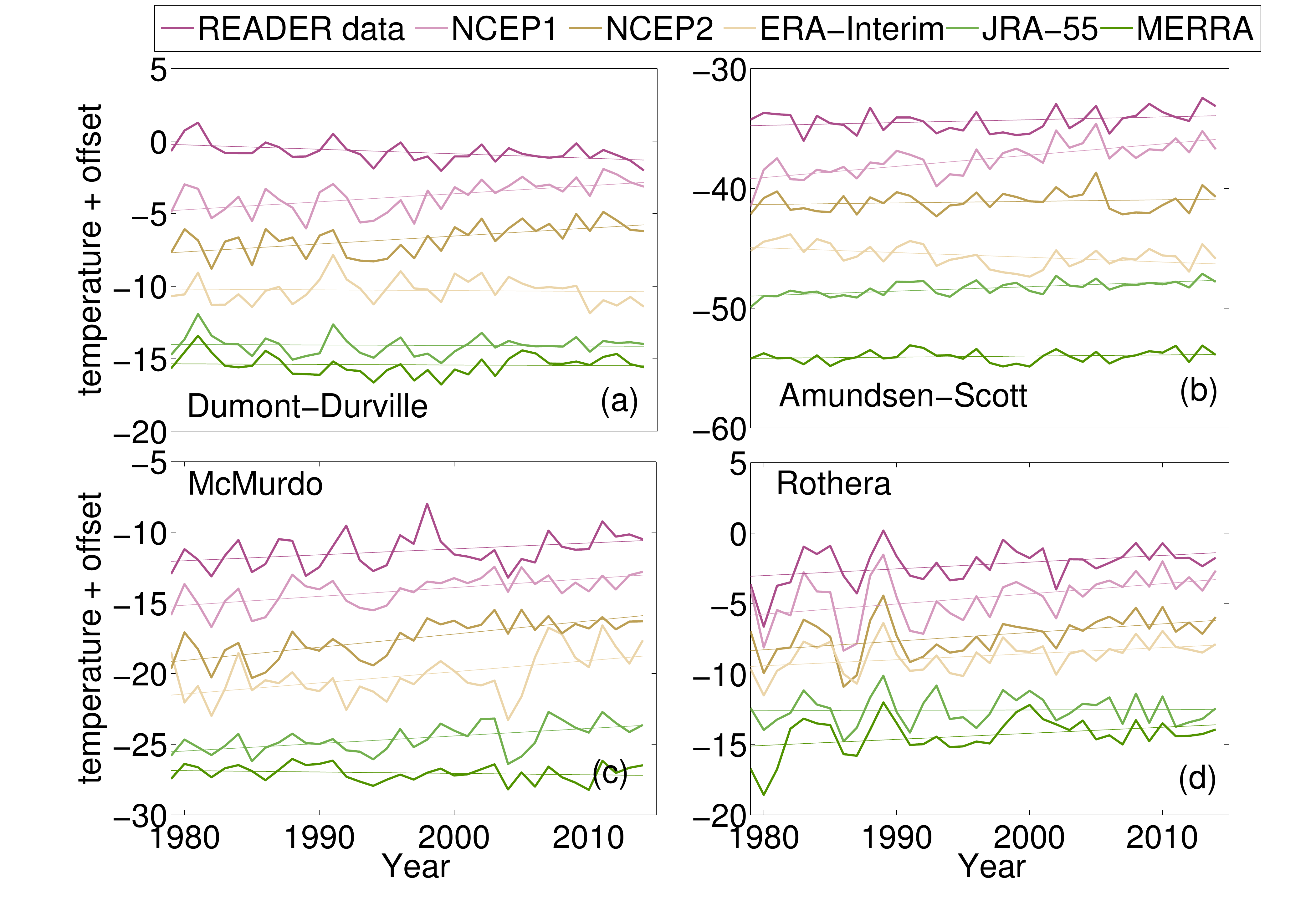}
  \caption{Comparison between annual observational $T_{2m}$ time series (blue) and 5 reanalysis projects (NCEP1 - black, NCEP2 - red, ERA-Interim - magenta, JRA55 - green and MERRA - cyan) for 4 stations, (a) Dumont-Durville, (b) Amundsen-Scott, (c) McMurdo, and (d) Rothera. The trend lines have been obtained by linear regression.}
  \label{Fig2}
\end{figure}

Figure~\ref{Fig3} shows the result of our quantitative analysis of $\Delta$ and the standard deviation $\sigma$ along the regression line. In the figures, Bellingshausen, Dumont-Durville and Amundsen-Scott are abbreviated by Bell, DD and AS, respectively. In the Antarctic Peninsula, the observational data show warming trends at Faraday and Rothera, while in Bellingshausen there is no obvious temperature change in the past 36 years. Almost all reanalysis datasets show warming patterns except JRA-55, and the warming trends are in reasonable agreement with the observations. JRA-55, however, shows almost no temperature change in the last 36 years in the Peninsula, which is inconsistent with the observational records. In West Antarctica, however, the ERA-Interim and JRA-55 reanalysis data are in good agreement with the observations, while the NCEP2 reanalysis exaggerates the warming trends considerably. In particular, NCEP2 overestimates the trends $\Delta$ at Byrd and McMurdo by a factor of 3.3 and 1.8, respectively. MERRA, on the other hand, show almost zero warming at Byrd and McMurdo. 

\begin{figure}[htpb]
    \centering
\includegraphics[width=\columnwidth,trim=50 50 50 50,clip=false]{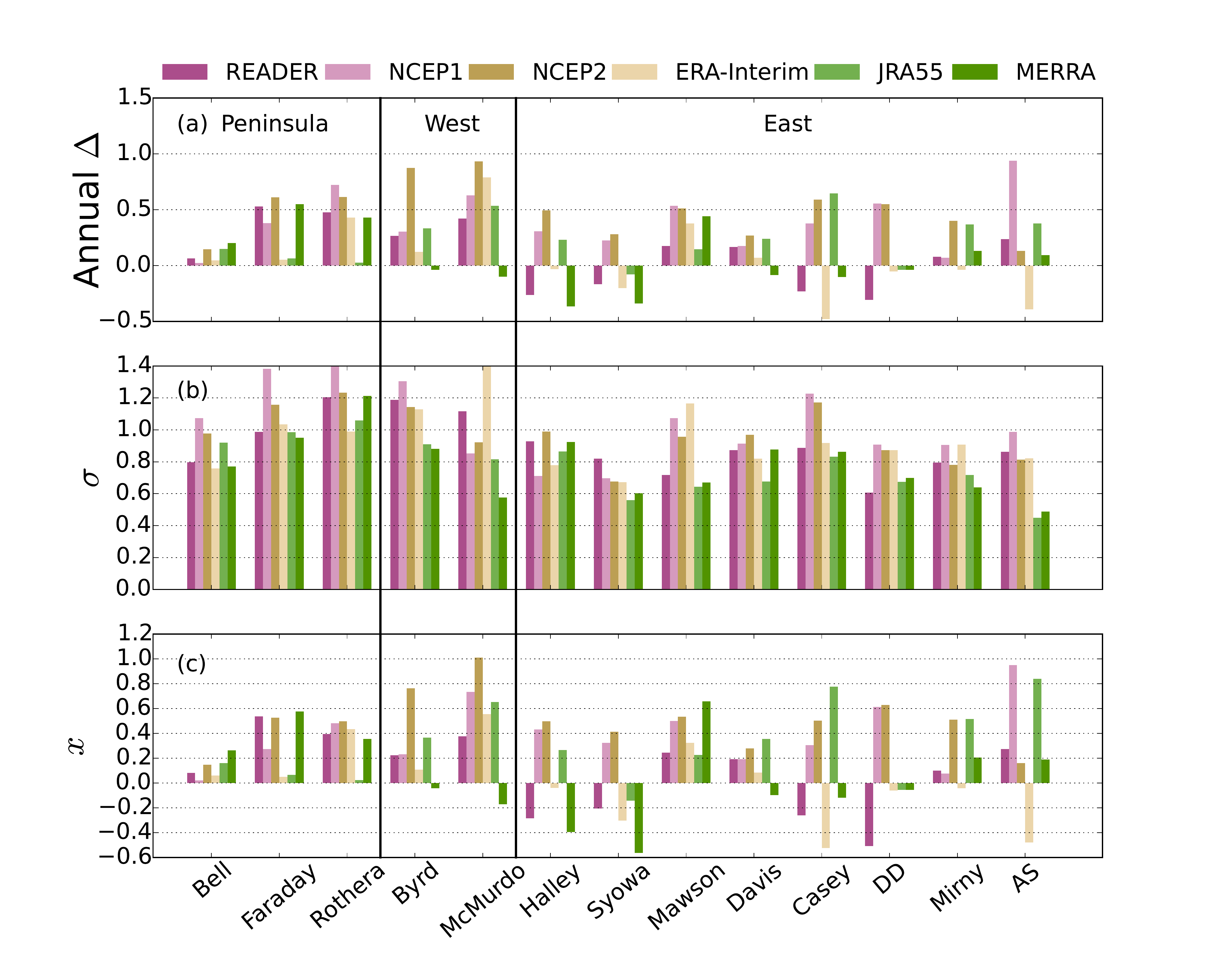}
  \caption{Comparison of (a) the absolute trend $\Delta$ per decade, (b) the standard deviation $\sigma$ around the regression line and (c) the relative trend $x=\Delta/\sigma$ between observation data and reanalysis datasets at different stations. Bellingshausen, Dumont-Durville and Amundsen-Scott are abbreviated to Bell, DD and AS, respectively. The units of $\Delta$ and $\sigma$ are degree Celsius.}
  \label{Fig3}
\end{figure}

In East Antarctica, Figure~\ref{Fig3} shows that the performance of NCEP1 and NCEP2 is not reliable, with considerable warming trends at Halley, Casey, Dumont-Durville and Amundsen-Scott. Moreover, the JRA-55 reanalysis data also exhibits a strong warming trend at Casey. In contrast, the observational data at Halley and Casey show cooling trends with $\Delta=-0.26\degree{\rm{C}}/\rm{decade}$ and $-0.23\degree{\rm{C}}/\rm{decade}$, respectively. For Amundsen-Scott, NCEP1 and JRA-55 exaggerate the warming trend $\Delta$ by a factor of 3.6 and 1.5, respectively. The performance of the ERA-Interim/MERRA reanalysis data in East Antarctica is considerably better. An exception for ERA-Interim perhaps is the negative trend at Amundsen-Scott ($\sim -0.4\degree{\rm{C}}/\rm{decade}$) that is inconsistent with the observation. {Indeed, recent careful investigations showed that ERA-Interim did not assimilate any observations for Amundsen-Scott before the start of 1986~\citep{Jones2014}.}

{The physical reason why West Antarctica as well as the Antarctic Peninsula exhibit a warming trend in the near surface temperature data has been addressed by previous studies. For example, the warming in West Antarctica during austral spring is partly due to the Pacific South American (PSA) mode, especially PSA-1 mode that is a wave-train extending from the tropics to the high Southern latitudes~\citep{Schneider2012}. Moreover, the warming in the  Peninsula may be due to the extratropical Rossby wave train associated with tropical Pacific sea surface temperature anomalies~\citep{Ding2011, Ding2013}{.}}

{For obtaining a quantitative measure for the goodness of the reanalysis data with respect to the observed annual temperature trends, we determined for each reanalysis data set the standard error $\delta_E$ with respect to the READER data set, }
{
\begin{equation}
	\delta_E=[\frac{1}{13}\sum_{i=1}^{13} (\Delta_i-\Delta_i^R)^2]^{1/2}, 
\end{equation}
and the typical observational trend $\delta_O$,
\begin{equation}
	\delta_O=[\frac{1}{13}\sum_{i=1}^{13} (\Delta_i^R)^2]^{1/2}.
\end{equation}}
Here, $\Delta_i$ is the temperature change of a specific reanalysis data set at  station $i$, $i=1,2,\cdots,13$, and $\Delta_i^R$ is the corresponding observational temperature change from the READER data set. Table I shows the dimensionless relative standard error  $\delta_r=\delta_E/\delta_O$ for all reanalysis data sets. 
We can obtain a threshold for $\delta_r$ by comparing the observational data with their shuffled counterpart. For the shuffled data, $\Delta_i\cong0$ for all stations, which leads to $\delta_r=1$. We consider a reanalysis data set as unsatisfying when the $\delta_r$ is above the threshold. When the signs of $\Delta_i$ have the same signs as the $\Delta_i^R$ of the observational data, then $\delta_r<1$. The lower is $\delta_r$, the better is the performance of the reanalysis data set.

{
As expected from the foregoing discussion, NCEP1, NCEP2, and JRA-55 show a quite poor performance, with  $\delta_r$ of the annual trend well above 1, $\delta_r=1.44, 1.66$, and 1.21, respectively. The best performance can be found in ERA-Interim and MERRA, with $\delta_r=0.96$ and 0.77, respectively. The very poor performance of NCEP1, NCEP2, and JRA-55 is a puzzle for us. It is unlikely that the inconsistency of the temperature change between observational data and reanalysis data is simply a manifestation of a low quantity of observations in this region, since many observational data have been assimilated after the year of 1979. The inconsistency may rather result from the data assimilation methods and model details, in particular from the biases in surface radiative flux that affects the surface temperature inversion. Note that JRA-55 does not show any temperature change in the Antarctic Peninsula in the past 36 years, which is inconsistent with the observational temperature change.}

{For the standard deviation $\sigma$ along the regression line, there is no dramatic difference between reanalysis and observation data except at Mawson and McMurdo, where the ERA-Interim reanalysis data slightly exaggerate the variability. Moreover, at Amundsen-Scott, $\sigma$ of JRA-55 and MERRA is 1/3 less than observational data. As a result, the relative temperature trend $x$ will show a very similar picture as the absolute trend $\Delta$ (i.e. Figure~\ref{Fig3} (a)). We have calculated our quality measure $\delta_r$ for the standard deviations. According to Table~\ref{Table1}, $\delta_r$ varies between 0.27 for NCEP1 and 0.18 for NCEP2, representing only small deviations between reanalysis and observational data. In this case, however,  unlike for the temperature trends, $\delta_r=1$ does not provide the threshold for shuffled data, since the shuffling method yields $\sigma$ unchanged.}

\subsection{Seasonal warming/cooling trends}

Next {we address the seasonal warming/cooling trends in Antarctica. The results are shown in Figure~\ref{Fig4}. In the Antarctic Peninsula, the reanalysis data sets show a reasonable performance, as for the annual trends discussed above. An exception is again JRA-55, which  during all  seasons shows less warming trends at Faraday and Rothera than the observational data. The less warming trend as reflected by JRA-55 challenges the reliability of its underlying model, since the observational data in this region were well assimilated after the year of 1979. Moreover, NCEP2 provides higher trends at Bellingshausen than the observational data during austral winter; ERA-Interim shows little evidence of warming trends at Faraday across all seasons.}

In West and East Antarctica where  the density of stations is low, the performance is much worse.  During austral winter, the warming trends of NCEP1 and NCEP2 at Byrd are exaggerated by a factor of 2.6 and 4.7, respectively (reaching 0.88\degree{C} and 1.58\degree{C} per decade). MERRA, however, does not show any temperature change during austral spring, which is inconsistent with the strong warming trend in the observational data.

In East Antarctica, NCEP1 and NCEP2 are unreliable in describing the temperature trends during non-summer seasons, where spurious warming trends are found at Halley, Casey and Amundsen-Scott. Moreover, JRA-55 shows considerable warming trends at Casey and Amundsen-Scott during austral non-summer seasons and austral summer, respectively. The observational data, in contrast, show either cooling (Halley and Casey) or modest warming (Amundsen-Scott).  ERA-Interim and MERRA perform considerably better, and show trends much closer to observation with slight deviations of ERA-Interim at Amundsen-Scott. 

{We have applied our accuracy measure $\delta_r$ to the seasonal trends in the reanalysis data sets. The results are listed in Table~\ref{Table1}. As for the annual trends, the accuracy of  NCEP2 is unacceptable, with $\delta_r$-values between 1.27 and 1.72. These high values can only occur when the reanalysis project produces warming while the observational data show cooling and vice versa. Surprisingly, NCEP1 is slightly better, with the best performance (0.93) in austral spring  and the worst performance in austral fall (1.94). The performance of JRA-55 is only slightly better,  again with the worst performance (1.37) in austral summer and the best performance (0.82) in austral spring. In contrast, ERA-Interim and MERRA have a reasonable performance, where our accuracy index is below 1 for all seasons except austral summer.}

\begin{figure*}[htpb]
    \centering
\includegraphics[width=\textwidth,trim=50 50 50 50,clip=false]{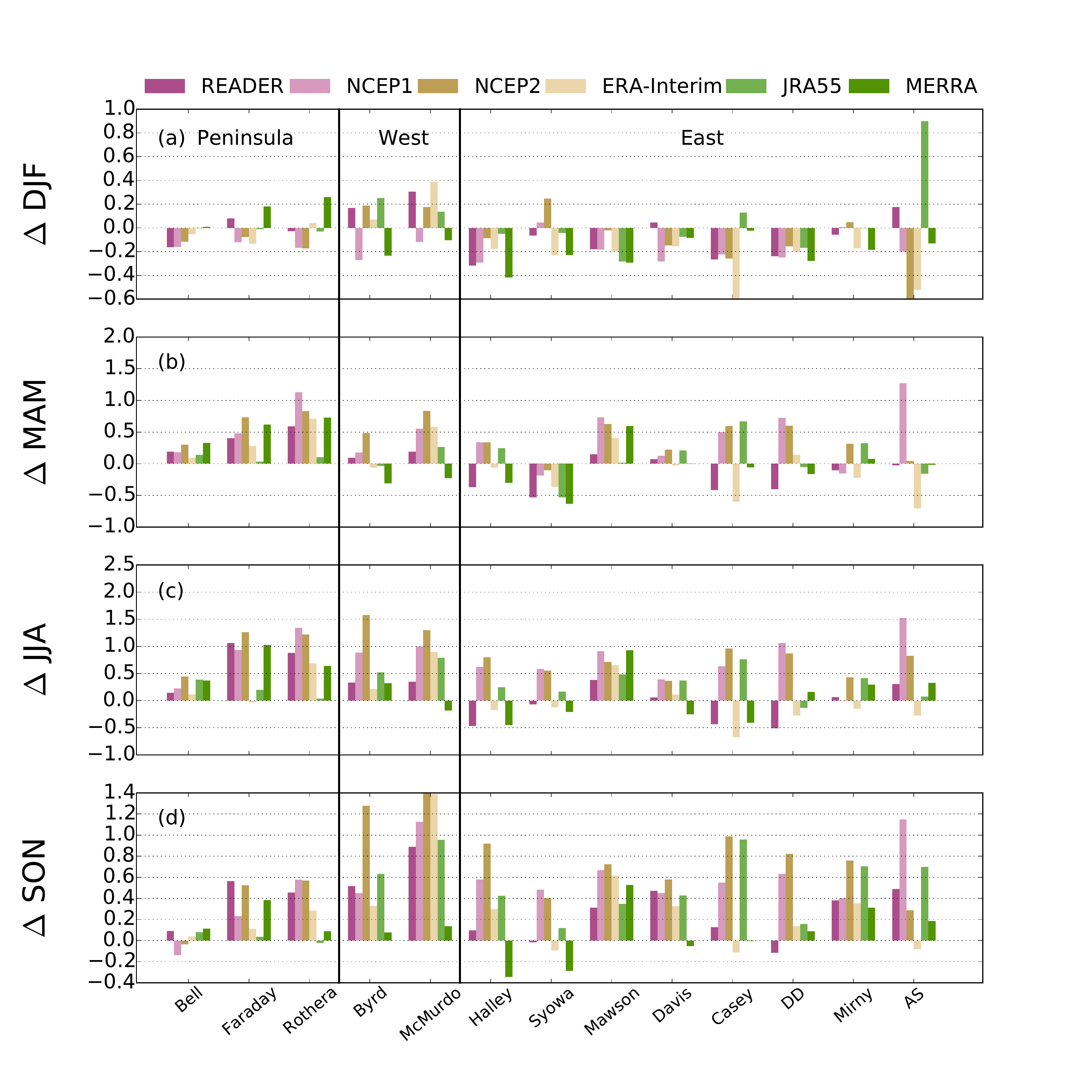}
  \caption{Same as Figure~\ref{Fig3}a, but for (a) austral summer (DJF), (b) austral autumn (MAM), (c) austral winter(JJA) and (d) austral spring (SON).}
  \label{Fig4}
\end{figure*}

\begin{figure*}[htbp]
    \centering
    \includegraphics[width=\textwidth,trim=50 50 50 50,clip=false]{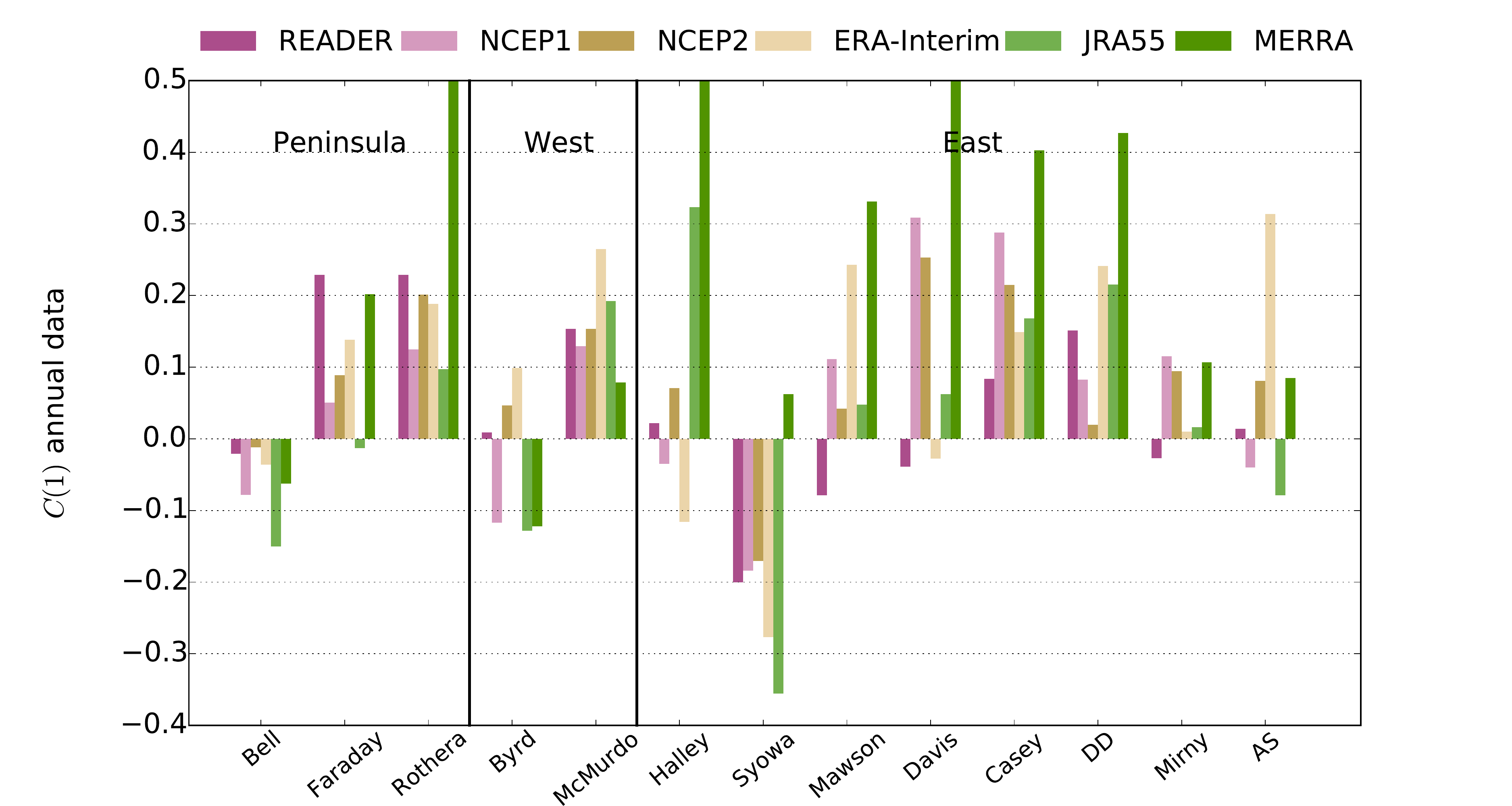}
\caption{Same as Figure~\ref{Fig3}, but for lag-1 autocorrelation $C(1)$ of detrended annual data.}
\label{Fig5}
\end{figure*}

\subsection{Persistence analysis: Lag-1 autocorrelation}

Next {we consider the linearly detrended lag-1 autocorrelation function $C(1)$ of the annual temperature data and the Hurst exponent $\alpha$. The reason why we consider $C(1)$ here is that it plays a central role in the conventional estimation of the statistical significance of a temperature trend as mentioned, see e.g., in the last IPCC report~\citep{IPCC}. The statistical significance is crucial when one wants to {assess} if an observed trend is likely to be of anthropogenic origin or not. The main assumption in the conventional treatment ~\citep{Santer2000} is that the detrended lag-$s$-autocorrelation function $C(s)$ of an annual temperature record is short-term persistent and decays, to a very good approximation, as 
\begin{equation}
	C(s)=C(1)^{-s},  s=0,1,2,\cdots.
\end{equation}}

{The second assumption is that also in short records $C(1)$ can be obtained sufficiently accurately. However, it has been shown~\citep{Lennartz2009} that this assumption is not fulfilled. The accuracy in the autocorrelation function depends on the record length $L$, and the results for $C(s)$ are only reliable for $s$ well below $L/50$. This means that in records of length $L$ below 50 even $C(1)$ cannot be calculated reasonably well.  This sheds doubts on the conventional significance analysis as long as short records are considered, even in the case that the record is short-term persistent.}

{Under the condition that the record is short-term persistent, the sign of $C(1)$ highly matters. When $C(1)$ is positive, the data are persistent. When $C(1)$ is negative, the data are anti-persistent. Under the assumption that the record is long-term persistent, $C(1)$ does not play an important role, and is replaced by the Hurst exponent $\alpha$. }

{Figure~\ref{Fig5} shows the linearly detrended lag-1 autocorrelation $C(1)$ obtained from the annual mean temperature data at  the different stations. The figure shows that for the same station, $C(1)$ fluctuates strongly in the different reanalysis projects (with a $\delta_r$-value even exceeding 2 for MERRA), fluctuating between positive and negative values. If the data were indeed short-term persistent as postulated in several articles \citep[e.g.,][]{Bromwich2013, Jones2014}, this would imply that for the same station, some reanalysis projects would produce short-term persistent temperature curves, while the others would produce anti-persistent curves. When observing the temperature data in {Figure} 2, however, such a difference cannot be recognized, since all curves for one station showed the same characteristic `mountain-valley' structure.}
For obtaining a threshold value for $\delta_r$, we consider again shuffled observational data where $C(1)=0$. In this case, as for the temperature trends considered above, $\delta_r=1$ is the threshold. Table 1 shows that except for NCEP2 where $\delta_r$ is slightly below 1, all $\delta_r$ values are above 1.
 
{Accordingly, the value of $C(1)$ can not describe sufficiently the structure of the temperature records and thus  $C(1)$ cannot represent a central quantity for the statistical significance of the data. Following \citet{Lennartz2009}, we believe that the large fluctuations in $C(1)$ between the different reanalysis projects  are mainly produced by the finite size effects in calculating $C(1)$, and not by different dynamical features in the reanalysis projects.}

\subsection{Persistence analysis: Long-term correlations and Hurst exponent}

{The previous subsection points to the possibility that a description of the temperature records as short-term persistent may not be appropriate. Indeed, when analyzing the observational READER data by the Detrended Fluctuation Analysis (DFA2) (see Methods Section for detailed mathematical descriptions) it has been shown recently \citep{Ludescher2015} that the form of the DFA2 fluctuation function $F(s)$ does not fit to the assumption of a short-term dependent process (where C(s) decays exponentially, as in  Eq. 4) but rather fits to the assumption of the existence of a long-term dependent process where the autocorrelation function $C(s)$ decays algebraically,}

\begin{equation}
	C(s) \propto (1-\gamma)s^{-\gamma},  0<\gamma<1, s>0.
\end{equation}
By definition, $C(0)=1$, and $\gamma=1$ describes white noise. The autocorrelation function $C(s)$, however, is not a good tool for detecting long-term correlations, since strong finite-size effects occur (see above), which strongly limit the validity range of $s$.
A more reliable tool is the  DFA method, where one calculates a fluctuation function $F(s)$. When $F$ has the form of a power law~\citep{Peng1994, Kantelhardt2001}, 
\begin{equation}
	F(s)\sim s^\alpha, 
\end{equation}
then this indicates a long-term persistent process, and the Hurst exponent $\alpha$ is related to the correlation exponent $\gamma$ by
\begin{equation}
	\alpha=1-\gamma/2. 
\end{equation}
For $\gamma$ between 0 and 1, $\alpha$ ranges between 1/2 and 1. Long-term persistence is not a feature of the Antarctic stations alone. Long-term persistence occurs all over the globe, in atmospheric temperatures as well as in sea-surface temperatures. Typical values for $\alpha$ are 0.65 for continental and coastline stations \citep{Koscielny1998, Malamud1999, Fraedrich2003,  Eichner2003,  Kiraly2006, Franzke2010, Franzke2012,Ludescher2015,Yuan2015}, 0.75 for island stations and 0.8 for sea-surface temperatures \citep{Monetti2003}. 
The error bars corresponding to the 95 percent confidence interval are $\pm 0.10$ for coast line and continental stations, $\pm 0.15$ for island stations, and
$\pm 0.10$ for sea surface temperatures.
Long-term persistence (long-term memory) does not only occur in temperature records, but is also known to characterize systems as diverse as river-flows \citep{Hurst1951, Mandelbrot1968, Tessier1996, Montanari2000,  Koutsoyiannis2006, Kantelhardt2006, Koscielny2006,  mudelsee2, Livina2003},  sea level heights \citep{Beretta2005, Dangendorf2014, Becker2014}, wind fields \citep{santhanam} and midlatitude cyclons  \citep{Blender2014}. Other examples include heartbeat intervals \citep{Bunde2000, Peng1993}, DNA sequences~\citep{Peng1992}, the volatility in financial markets \citep{Bunde2012}, and the arrangement of rare words in literary texts \citep{Ebeling, Altmann}. 

{We have performed a DFA2 analysis and determined $F(s)$ for all temperature records considered here. Typical result for two stations, Rothera and Dumont-Durville, are shown in Figure~\ref{Fig6}. The figure shows that all fluctuation functions display the same perfect power-law behavior, from which we can easily extract the Hurst exponent as slope in the double-logarithmic plots. The result for the Hurst exponents of all records is shown in Figure~\ref{Fig7}.  One can see that the Hurst exponents for the observable data found by us in Antarctica are within the ranges expected for continental, coastline and sea surface temperatures discussed above.
The figure shows that the Hurst exponents of the 5 reanalysis projects and the READER data are very close to each other, revealing that all the reanalysis projects are able to reproduce the long-term persistent nature of the instrumental data.  Indeed, this may be expected from {Figure} 2, since the reanalysis data show the same characteristic `mountain-valley' structure  generated by the long-term memory in the data as the instrumental data. Our accuracy measure gives the same very low value for $\delta_r$ for all reanalysis projects considered (see Table~\ref{Table1}).} A threshold value for $\delta_r$ can be obtained, as for the temperature trends and $C(1)$, from shuffling the READER data. When shuffling, the long-term persistence vanishes, and the Hurst exponents become 1/2. This leads to the threshold value 0.31, which is above the $\delta_r$ values for all reanalysis data sets.}

\begin{figure*}[htbp]
    \centering
    \includegraphics[width=0.9\textwidth,trim=50 50 50 50,clip=false]{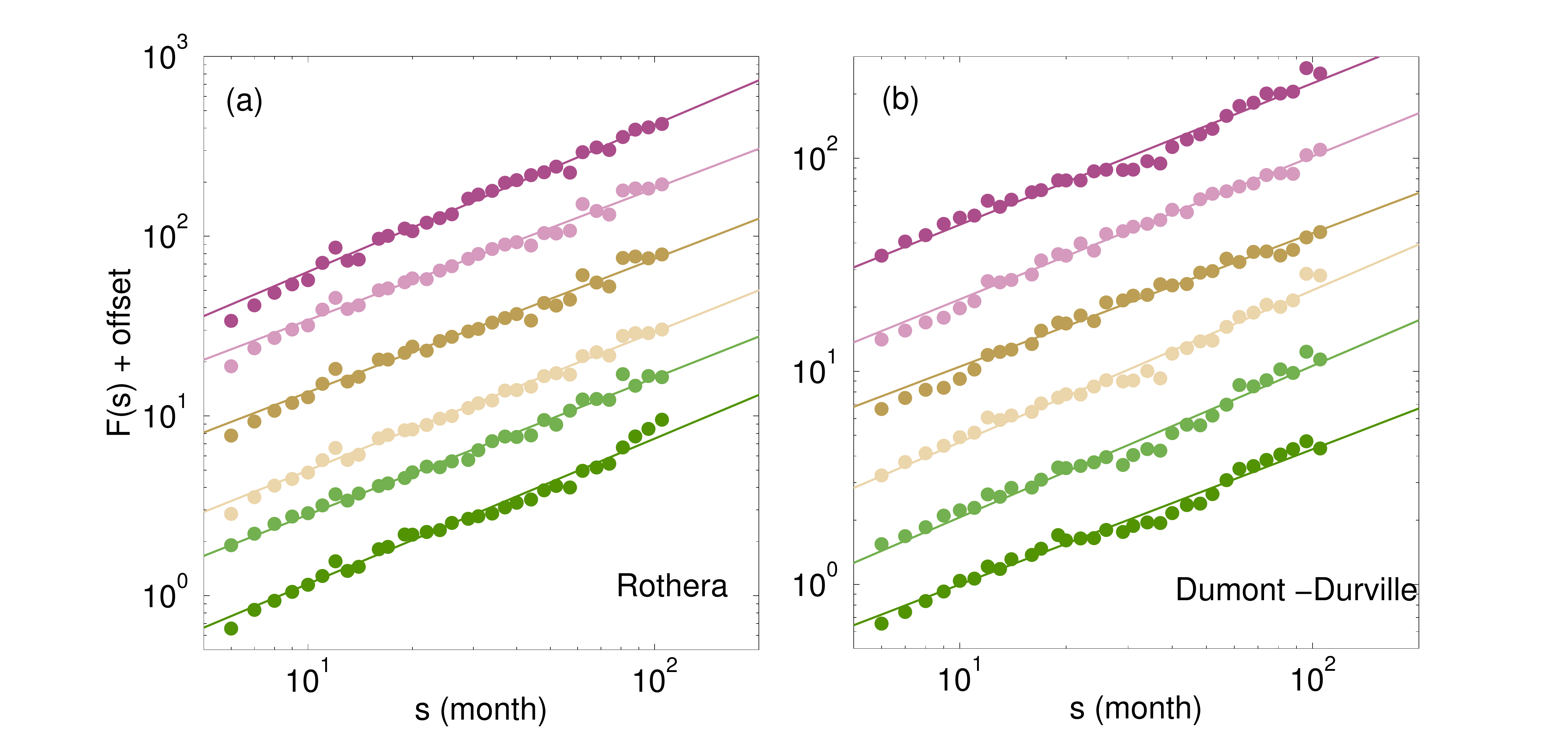}
\caption{Persistence analysis (DFA2). The figure shows the DFA2 fluctuation functions {$F(s)$} for two representative stations (a) Rothera, and (b) Dumont-Durville. The fluctuation functions can be well approximated, for $s$ above 10, by a power-law dependence.}
\label{Fig6}
\end{figure*}

\begin{figure*}[htbp]
    \centering
    \includegraphics[width=\textwidth,trim=50 50 50 50,clip=false]{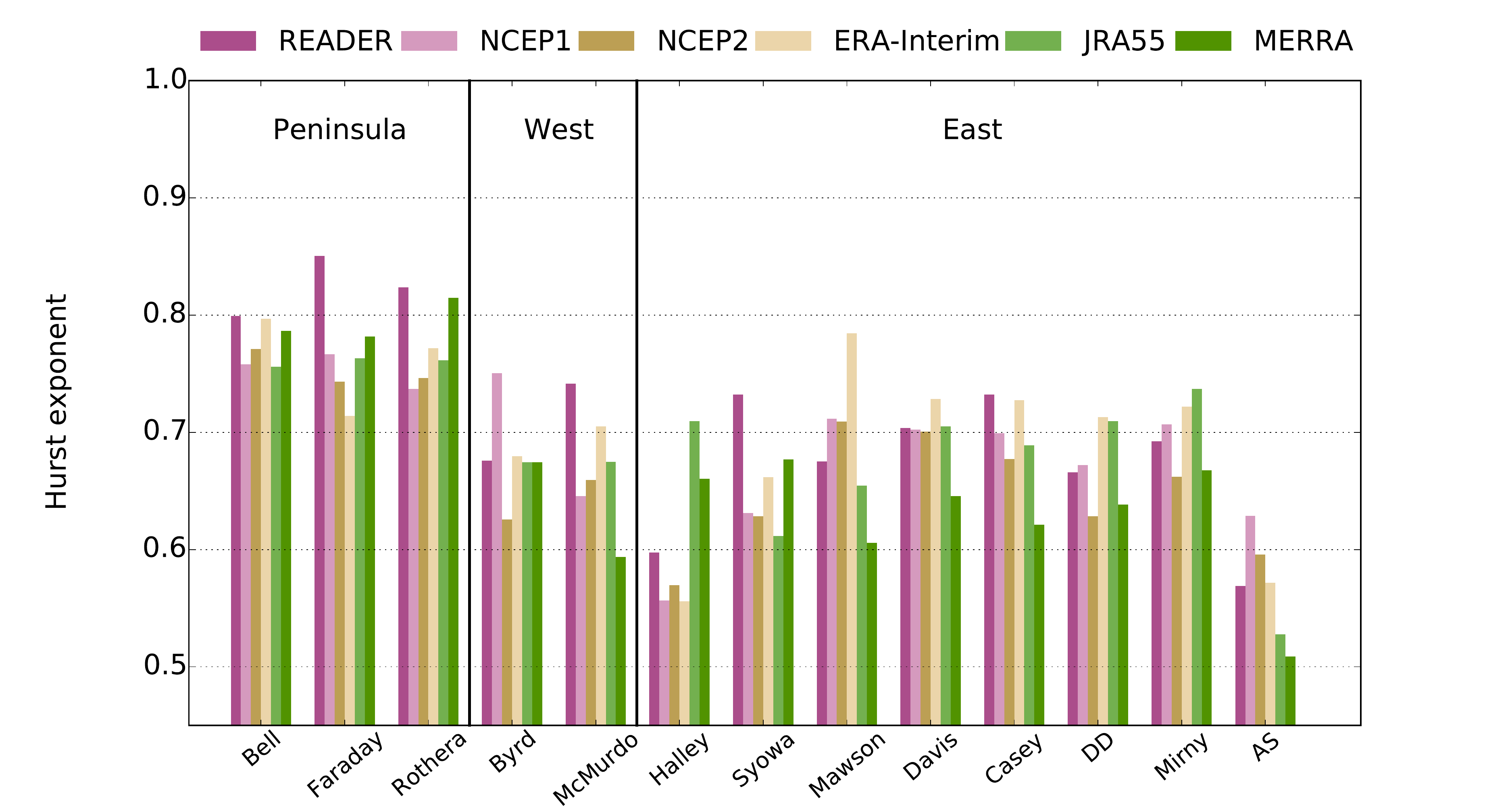}
\caption{Same as Figure~\ref{Fig3}, but for the DFA2 Hurst exponent $\alpha$ of the monthly temperature anomalies.}
\label{Fig7}
\end{figure*}

\subsection{Significance of trends}

We have described, in the Methods Section, how the significance $S$ of a trend can be obtained for (a) a short-term persistent process characterized by $C(1)$ and (b) a long-term persistent process characterized by the Hurst exponent $\alpha$. Despite the evidence that the Antarctic temperature data are long-term persistent ({see Figure 6, \citet{Franzke2010,Franzke2012}, \citet{Bunde2014}, \citet{Ludescher2015} and \citet{Yuan2015}}), we discuss the significance level of warming/cooling trends also for the hypothetical case of  short-term persistence.

\begin{figure*}[htbp]
    \centering
    \includegraphics[width=\textwidth,trim=50 50 50 50,clip=false]{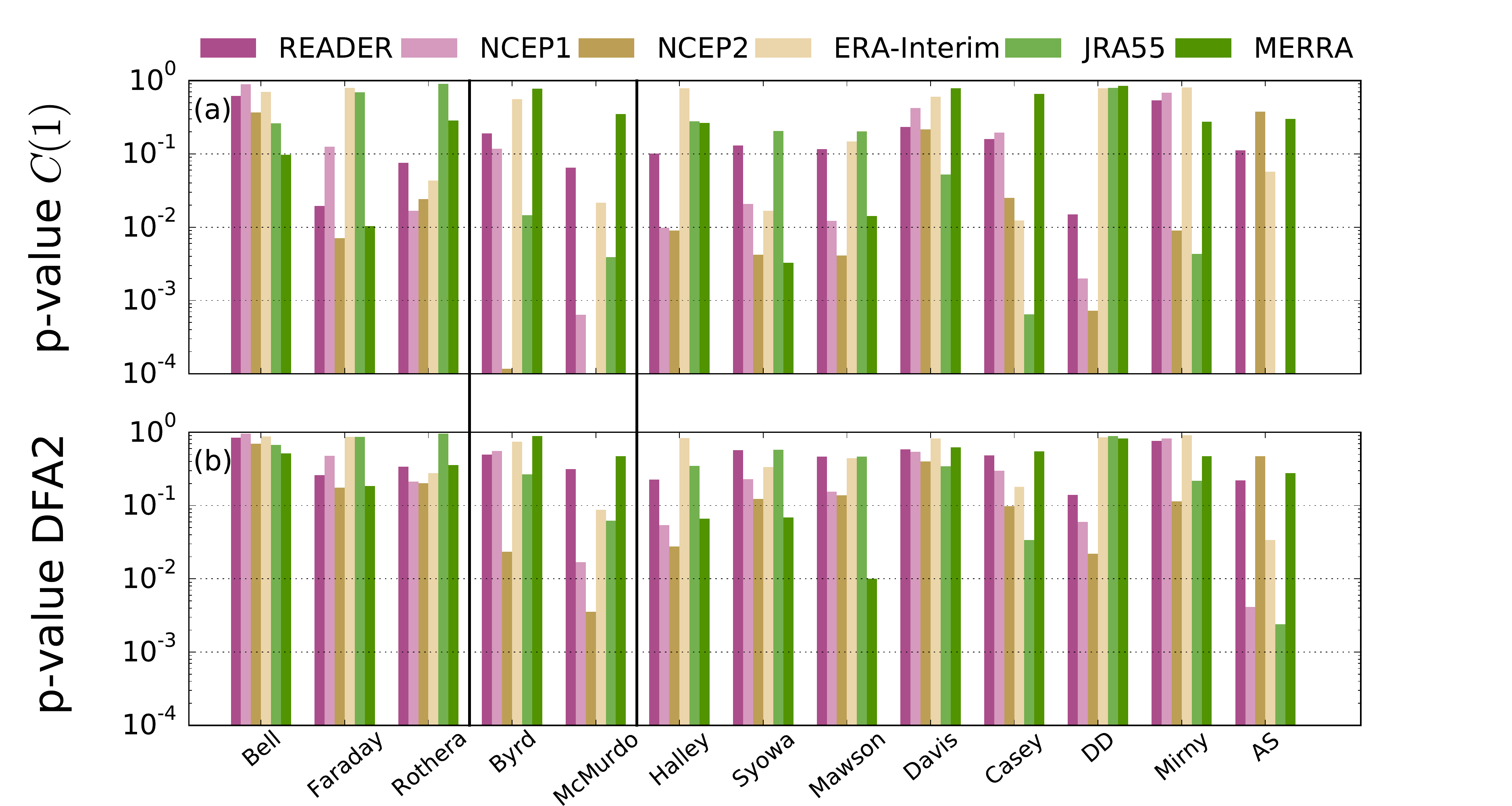}
\caption{Same as Figure~\ref{Fig3}, but for the $p$-value of the relative trend $x$, when assuming (a) an AR(1) process and (b) a long-term persistent process.}
\label{Fig8}
\end{figure*}

The results for the $p$-values ($p\equiv 1-S$) are shown in Figure~\ref{Fig8}: (a) under the hypothesis that the data are short-term persistent; and (b) for the assumption that the data are long-term persistent. We would like to emphasize that this result is only valid for the considered time window of {36-yr}, since our purpose is to assess the skills of reanalysis data sets. For most observational records the time window is indeed longer. Since the statistical significance of a trend increases with increasing length, the significance of the observational trend in the full time window might be larger than for the considered time window (see Methods Section and {\citet{Tamazian2015}}).

\begin{table}
\caption{The dimensionless relative standard error $\delta_r = \delta_E/\delta_O$ for annual trend, seasonal trend, standard deviation around the regression line, Hurst exponent $\alpha$, as well as the lag-1 autocorrelation function $C(1)$. The red $\delta_r$ are the ones with value smaller than the 5\% significance level of the null model.}
\centering
\begin{tabular}{l l l l l l l l l}
\hline
& \textbf{Annual} & \textbf{JJA} & \textbf{SON} &\textbf{DJF} & \textbf{MAM} & \textbf{$\sigma$} & \textbf{$\alpha$} & \textbf{$C(1)$}\\
\hline
\hline
NCEP1 & 1.44 & 1.62 &\textcolor{red}{0.93} & 1.24 & 1.94 &\textcolor{red}{0.27} & \textcolor{red}{0.08} &1.23\\
NCEP2 & 1.66 & 1.72 & 1.27 & 1.47& 1.68&\textcolor{red}{0.18} &\textcolor{red}{0.08}&0.93\\
ERA-I & \textcolor{red}{0.96} & \textcolor{red}{0.85} & \textcolor{red}{0.70} & 1.32& \textcolor{red}{0.94}&\textcolor{red}{0.21} &\textcolor{red}{0.08}&1.12\\
JRA55 & 1.21& 1.16 & \textcolor{red}{0.82} & 1.37 & 1.30&\textcolor{red}{0.21} &\textcolor{red}{0.09}&1.05\\
MERRA &\textcolor{red}{0.77} & \textcolor{red}{0.65} & \textcolor{red}{0.85} & \textcolor{red}{1.23} & \textcolor{red}{0.79}& \textcolor{red}{0.24}&\textcolor{red}{0.09}&2.08\\
\hline
\end{tabular}
\label{Table1}
\end{table}

Under hypothesis (a), only the observational records at Faraday and Dumont-Durville have $p$-values below 0.05 and are thus statistically significant. In contrast, NCEP1, NCEP2, and JRA-55 produce highly significant trends at several stations in West and East Antarctica. As expected from Figure~\ref{Fig3}, the ERA-Interim and MERRA data sets produce more realistic trend significances.  There are only few exceptions, for example ERA-Interim at Amundsen-Scott implies a significant cooling trend that is not observed in reality; MERRA exhibits a significant warming trend at Mawson. 

Under the more realistic assumption (b), all observational records have $p$-values even above 0.1, and thus the warming/cooling trends are not statistically significant in the considered time window. It is remarkable that despite all their differences, in the Antarctic Peninsula, the reanalysis data produce a comparable significance level, {revealing that 
in the time period after 1979, the warming in the Peninsula is not significant.} In the rest of Antarctica, however, NCEP1 and NCEP2 again show significant spurious warming trends at many stations. At Halley NCEP2 produces highly significant trends with $p$ below $0.05$. In contrast, the observational data have $p$ values above 0.1, and the contrast could hardly be higher. Again, JRA-55 shows significant warming at Casey and Amundsen-Scott ($p\leq0.05$). ERA-Interim (except at Amundsen-Scott) and MERRA (except at Mawson) perform much better.  

\section{Conclusions}

{In this paper, we assessed, for the first time, the warming trend and its significance as well as the persistence properties of five widely used global reanalysis datasets (NCEP1, NCEP2, ERA-Interim, JRA-55 and MERRA) in Antarctica. We considered the time period from January 1979 to December 2014, when modern satellite data were assimilated into the reanalysis datasets. We compared the reanalysis datasets with the longest observational $T_{2m}$ data from staffed observation stations across Antarctica. In our performance test, we first compared the absolute trends $\Delta$ and the standard deviation along the regression line $\sigma$. Then we considered the seasonal  warming trends. Finally, we studied the persistence properties (lag-1 autocorrelation $C(1)$ and Hurst exponent $\alpha$) as well as the significance level ($p$-value) of the relative trends.}

 {We found that all 5 reanalysis data sets were able to reproduce nicely  the long-term persistence in the {instrumental} data, with Hurst exponents quite close to each other. Also, the standard deviations along the regression lines are in very good agreement between reanalysis and instrumental data sets.  In contrast, $C(1)$, which is needed as input for the conventional significance analysis shows fully erratic behavior, and the observational warming/cooling trends in East  and West Antarctica (where the data are sparse) are not reproduced well by the reanalysis data sets. The worst performance showed NCEP1, NCEP2, and JRA-55  with spurious warming trends even in those parts of East Antarctica where cooling has been observed. In the Peninsula where the station density is quite high, the performance of the reanalysis data is considerably better. Under the assumption of a long-term persistent process all reanalysis data sets as well as the observational data show  that  in the considered time period since 1979 the warming in the Peninsula is not significant with  $p$ values well above 0.1{.}}
 
{ Overall, ERA-Interim and MERRA showed the best performance when testing the local temperature data in Antarctica.
Further work is needed in order to see if the data sets are also better in reproducing the known teleconnections between locations on different parts of the globe{.}}






\section{Methods}

In earlier studies on Antarctic warming ~\citep[e.g.,][]{Steig2009, Ding2011, Bromwich2013, Nicolas2014,Jones2014}, it has been assumed that the linearly detrended annual mean temperatures  $\Theta_i$ can be described by a first-order autoregressive process (AR(1)), which is defined as
\begin{equation}
	\Theta_{i+1} = r_1\Theta_i+\eta_i, i=1,2,\cdots,L-1,
\end{equation}
where $r_1$ is the AR(1) persistence parameter lying between -1 and 1, and $\eta_i$ is  Gaussian white noise. The data are  persistent if $r_1 > 0$ and anti-persistent if $r_1<0$. For $r_1=0$, they represent Gaussian white noise. The autocorrelation function of the linearly detrended record is defined as
\begin{equation}
	C(s)=\langle \Theta_{i}\Theta_{i+s} \rangle=\frac{\frac{1}{L-s}\sum_{i=1}^{L-s}\Theta_i\Theta_{i+s}}{\frac{1}{L}\sum_{i=1}^L \Theta_i^2},
\end{equation}
where $s$ represents the time lag. It can be shown that for AR(1) processes, in the limit
of $L\rightarrow \infty$, $C(s)$ decays exponentially, $C(s) = r_{1}^{-s} \equiv \rm{exp}(-s|\rm ln r_1|) \equiv \rm{exp}(-s/s_x)$, where $r_1$ is identical to the lag-1 autocorrelation $C(1)$ and $s_x=1/\vert\ln r_1\vert$ is the persistence time. Accordingly, for an AR(1) process, $C(1)$ is the relevant quantity, as long as $L$ is sufficiently large (which according to~\citet{Bunde2014} and~\citet{Ludescher2015}, however, is not the case for the annual Antarctic records). When assuming the relation $r_1=C(1)$, the significance $S$ of the relative trend $x$  can be written as~\citep{Santer2000,Tamazian2015}

\begin{equation}
S(x;L)=\frac{2x}{a}\cdot\frac{\Gamma\left(\frac{1}{2}\left(l(L)+1\right)\right)}{\sqrt{\pi l(L)}\Gamma(\frac{l(L)}{2})} \times 
\leftidx{_2}F_{1}\left(\frac{1}{2},\frac{1}{2}(l(L)+1);\frac{3}{2};-\frac{(x/a)^2}{l(L)}\right).
\label{Sign}
\end{equation}
Here, $\Gamma$ denotes the Gamma function, and $\leftidx{_2}F_{1}$ is the hypergeometric function~\citep{Tamazian2015}. The effective length $l(L)$ is $l(L) = L\frac{1-C(1)}{1+C(1)}-2$, and  the scaling parameter $a$ is given by
\begin{equation}
a=\frac{\sqrt{12}(L-1)}{\sqrt{L^2+2}}\cdot \frac{1}{\sqrt{l(L)}}.
\end{equation}

However, it has been argued recently that the temperature records in Antarctica~\citep{Bunde2014,Tamazian2015,Ludescher2015,Yuan2015}, as well as in other places of the globe are not short-term, but long-term persistent~\citep[e.g.,][]{Koscielny1998,Eichner2003, Fraedrich2003}. In a long-term persistent stationary process, the autocorrelation function decays algebraically as $C(s) = (1-\gamma)s^{-\gamma}$, where the correlation exponent $\gamma$ is between 0 and 1. For detecting long-term memory one usually does not consider $C(s)$ because it exhibits strong finite size effects, but the detrended fluctuation analysis called DFA~\citep{Peng1994,Kantelhardt2001}. The current standard method for detecting long-term memory in climate records is DFA2~\citep{Kantelhardt2001}, which is a modification of DFA~\citep{Peng1994} and eliminates linear external trends.

In DFA2, one considers the monthly temperature anomalies $\tilde T_i, i=1,2,\cdots N$ where the seasonal trend has been subtracted from the data.
Then one determines the ``profile"  $Y(i) = \sum_{k=1}^i \tilde T_k,i=1, 2, \cdots, N$. To obtain the fluctuations of $Y$ on time scales of length $s$ one
divides the record into $N_s$ equal non-overlapping segments of fixed length $s$. In each segment $\nu=1, 2, \cdots, N_s$, one  calculates the best quadratic fit $f_{\nu}(i)$ of the profile
and determines the variance
\begin{equation}
F_s^2(\nu) = \frac{1}{s} \sum_{j=1}^{s}\left[Y((\nu-1)s+j)-f_\nu((\nu-1)s+j))\right]^2
\end{equation}
around the fit. Averaging  $F_s^2(\nu)$ over all $N_s$ segments then gives the square of the desired fluctuation function $F(s)$.

The behavior of $F(s)$ depends on how the autocorrelation function $C(s)$ decays with $s$. If $C(s)$ decays exponentially, then $F(s)$ increases as $F(s)\propto s^{1/2}$ for $s$ well above the persistence time $s_x$. In the case of long-term correlations, 
$F(s)$ increases as
\begin{equation}
F(s)\propto s^{\alpha} 
\end{equation}
with the Hurst exponent $\alpha = 1-\gamma/2 > 0.5$ (the higher the value of $\alpha$, the stronger the long-term memory). The results for $F(s)$ are reliable for $s$ between 7 and $N/4$~\citep{Kantelhardt2001}. 

The Detrended Fluctuation Analysis (DFA2) has been applied to a large number of temperature records all over the globe \cite[e.g.,][] { Koscielny1998,  Fraedrich2003, Eichner2003,  Lovejoybuch} revealing the long-term persistent nature of temperature records. It has been shown in \cite{Vyushin2004} that a major cause of the long-term memory is volcanic forcing.

DFA2 has been also applied to the monthly Antarctic records considered here in \cite{Bunde2014}, \cite{Bromwich2014}, \cite{Ludescher2015}, \cite{Yuan2015}, and \cite{Tamazian2015}. It has been shown explicitly in \cite{Bunde2014} and \cite{Ludescher2015} that for each record, the fluctuation function $F(s)$ agreed with the fluctuation function of long-term correlated surrogate data with the same length and Hurst exponent $\alpha$.

It has been shown in \cite{Rybski2008} when considering millennium runs that the Hurst exponent was the same for monthly, annual and bi-annual data, i.e. did not depend on the length of the averaged region in the time series. For a meaningful DFA analysis, the length of the record should not be smaller than 400 data points, and thus for the Antarctic stations we can consider only monthly data. 

{ When a record is fully characterized by a certain Hurst exponent $\alpha$, the significance $S$ of a relative trend $x$ depends only on $\alpha$ and the record length $N$. It has been shown recently by \cite{Tamazian2015} that $S(x,N)$ also follows Eq.~\ref{Sign}, but with different parameters $a$ and $l$. These parameters depend on $\alpha$ and $N$ and have been tabulated (for $0.5\le \alpha\le 1.5$) and $N\ge 400$ in \cite{Tamazian2015}. Accordingly, for given $\alpha$ and $N$, the trend significance can be obtained straightforwardly from \cite{Tamazian2015}, making the trend estimation as easy as for short-term persistent processes. Analytic approximate formulas for the significance of the trends in long-term persistent records can be found in \cite{Lennartz2009,Lennartz2011}.}

\begin{acknowledgments}
The authors would like to acknowledge the support of the LINC project (no. 289447) funded by the EC's Marie-Curie ITN program
(FP7-PEOPLE-2011-ITN) and the Israel Science Foundation for financial support.
The data sets used in this work can be accessed through the following sources.

\noindent NCEP1 is provided by the National Oceanic and Atmospheric Administration (NOAA), from website ``http://www.esrl.noaa.gov/psd/data/gridded/data.ncep.reanalysis.html''.

\noindent NCEP2 is provided by NOAA, from website ``http://www.esrl.noaa.gov/psd/data/gridded/ \\  data.ncep.reanalysis2.html''.

\noindent ECMWF ERA-Interim is provided by the European Centre for Medium-Range Weather Forecasts (ECMWF), from website ``http://www.ecmwf.int/en/research/climate-reanalysis/ \\ era-interim''.

\noindent JRA-55 is provided by the Japan Meteorological Agency (JMA), from website ``http://jra.kishou.go.jp/JRA-55/index\_en.html''.

\noindent MERRA is provided by the Global Modelling and Assimilation Office (GMAO), National Aeronautics and Space Administration (NASA), from website ``https://gmao.gsfc.nasa.gov/ \\ reanalysis/MERRA/''.

\noindent READER is provided by the Scientific Committee on Antarctic Research (SCAR), from website ``https://legacy.bas.ac.uk/met/READER/data.html''.

\end{acknowledgments}



\end{article}

\end{document}